\documentstyle[aps,pre]{revtex}

\begin{document}
\draft
\title{Decay of Classical Chaotic Systems --- \\
the Case of the Bunimovich Stadium}

\author{H. Alt$^{(1)}$, H.-D. Gr\"af$^{(1)}$,
 H.L. Harney$^{(2)}$,
  R. Hofferbert$^{(1)}$,
  H. Rehfeld$^{(1)}$, A. Richter$^{(1)}$ and\\
 P. Schardt$^{(1)}$}
\address {$^{(1)}$
   Institut f\"ur Kernphysik, Technische Hochschule Darmstadt,\\
   D-64289 Darmstadt, Germany\\
 $^{(2)}$
    Max-Planck-Institut f\"ur Kernphysik,\\
   D-69029 Heidelberg, Germany
}

\date{\today}
\maketitle
\begin{abstract}
The escape of an ensemble of particles from the Bunimovich stadium via a small
hole has been studied numerically. The decay probability starts out
exponentially but has an algebraic tail. The weight of the algebraic decay
tends to zero for vanishing hole size. This behaviour is explained by the slow
transport of the particles close to the marginally stable bouncing ball orbits.
It is contrasted with the decay function of the corresponding quantum system.
\end{abstract}
\twocolumn
\narrowtext
\noindent
\section{Introduction}
\label{sec1}
The decay of a quantum mechanical system whose states may be treated
statistically is --- on the average over the states --- not always exponential.
It is algebraic, if only a few decay channels are open.
This has been demonstrated in \cite{Alt1}.
There, the quantum mechanical system was simulated experimentally by a
quasi-two-dimensional microwave cavity. The classical analog to this is the
motion of a particle in a billiard shaped like a stadium
which was  shown by Bunimovich
\cite{Bunim1} to be fully chaotic and especially ergodic. By the present note,
we want to show that the algebraic decay observed in \cite{Alt1} is a quantum
mechanical feature that has no classical counterpart. To this end we have
studied the escape probability $P(t)$ via a small hole out of the classical
Bunimovich stadium and
find $P(t)$ ``almost exponential''.
This means: Every $P(t)$ has an algebraic tail for
large $t$, but the decay functions approach exponential behaviour with
decreasing size of the hole.

The algebraic asymptotics of the decay function $P(t)$ can be viewed as an
example of the ``third type of decay'' mentioned in the introduction of
\cite{HBS}, where ``particles initialized in a chaotic region can stick
a long time to the vicinity of the boundary of a regular region''.
Although there is no region of regular motion in the phase space of the
billiard there exist marginally stable orbits that cause this ``sticking''. A
schematic model will let us --- semiquantitatively --- understand why the decay
is nevertheless almost exponential.

The results of the numerical simulation are presented in Sec.~\ref{sec2}.
They are contrasted with the decay of the corresponding quantum system
in Sec.~\ref{sec3}. The simple model illustrating the interplay
between exponential and algebraic decay of the classical system is described
in Sec.~\ref{sec4}. Its predictions are compared with the present results
in Sec.~\ref{sec5}.

\section{Numerical experiment}
\label{sec2}
As in \cite{Alt1}, a quarter of the Bunimovich stadium has been considered in
order to remove symmetry. Its shape is sketched in the insert of
Fig.~\ref{fig1}. All lengths are given in
units of the radius of the circular
part of the boundary. The shape parameter $\gamma$ (see Fig.~\ref{fig1}) was
$\gamma=1.8$. An escape hole of size $\Delta$ has been assumed
--- as indicated in Fig.~\ref{fig1} --- in the
upper half of the small straight piece of the boundary. For
an ensemble of $10^6$ particles,
initial conditions were chosen at random. The distribution was a constant times
$dxdyd\phi$, where $x$ and $y$ are Cartesian coordinates of the position
inside the stadium and $\phi$ is an angle characterising the direction
of motion.
The orbit of every particle was followed numerically until it escaped via the
hole out of the stadium. The orbits were, however, not followed beyond $10^5$
collisions between the particle and the boundary. For a hole
of size $\Delta=0.05$ the result of this numerical experiment is given in
Fig.~\ref{fig1}
by the histogram of the probability density $P(L)$ for the particle to escape
after an orbit of length $L$. This will be called data in the sequel.
Since the velocity of the particle has constant modulus, we identify $L$
with time.

One finds $P(L)$ exponential for sufficiently small $L$. Later $P(L)$ turns
into a more slowly decaying function. This behaviour is typical and it occurs
wherever the hole is positioned.
The probability density $P(L)$ can be
represented by the combination of exponential and algebraic functions
\begin{equation}
P(L)=E \lambda \exp{(-\lambda L)}+A\alpha(\beta-1)(1+\alpha L)^{-
\beta},
\label{gl21}
\end{equation}
where $E+A=1$ for a normalized $P$. Then $A$ measures the weight of the
algebraic term. The full curve in Fig.~\ref{fig1} is a fit
to the data. The parameters
$\alpha,\beta,\lambda$ and $A$ were searched for. This procedure was
followed in 15 ``experiments'' with holes in the range
of \mbox{$0.25\ge\Delta\ge
0.0025$}. The results are partly reproduced in Tab.~\ref{tab1}
together with the values of the normalized $\chi^2$. Although only
the $\gamma=1.8$ stadium is discussed and analyzed below, we display in
Tab.~\ref{tab1} also some results pertaining to the $\gamma=1$ stadium in
order to show that it behaves quite similarly.

Table~\ref{tab1} and Fig.~\ref{fig2} show that the weight $A$ of the
algebraic decay approaches zero for $\Delta\to0$, i.e. the decay of
the stadium becomes exponential in the limit of a vanishing size of the hole.
We call this behaviour ``almost exponential'' decay.

In the limit of $\Delta\to0$ the decay constant approaches
--- see Tab.~\ref{tab1} --- the value
\begin{equation}
\lambda_0=\frac{\Delta}{\pi A_c}=\frac{4\Delta}{\pi(\pi+4\gamma)}
\label{gl22}
\end{equation}
given in \cite{BaBe}. Here, $A_c$ is the area of the
billiard; the momentum of the particle has been set equal to unity.
Equation~(\ref{gl22}) can be derived from ergodicity: Every point in
phase space should be visited with equal probability independent of the
elapsed time. From this one infers \cite{BaBe} that the decay should be
exponential with the decay constant $\lambda_0$.

We note that the dynamics
of the stadium may also be described as a mapping which generates --- from
given coordinates of collision with the boundary --- the coordinates of the
next collision. If one uses the number of mappings applied to the initial
distribution as time variable one again infers exponential decay, since
the motion on the boundary is ergodic, too. The decay constant $\nu_0$ is then
simply the ratio of the size of the hole and the perimeter of the billiard,
\begin{equation}
\nu_0=\frac{2\Delta}{\pi+2\gamma}.
\label{gl23}
\end{equation}
Here, we consider as boundary of the quarter stadium only the piece that it
shares with the original full stadium. We disregard the ``lines of cut''.
Similarly  a ``collision with the boundary'' is defined as a collision with
the piece of the original boundary (one could as well include the full
boundary of the quarter stadium into these definitions but the present choice
is more convenient). The ratio
\begin{equation}
\frac{\nu_0}{\lambda_0}=\frac{\pi(\pi+4\gamma)}{2(\pi+2\gamma)}
\label{gl24}
\end{equation}
then is the mean free path of the particle between two collisions with the
boundary. We have verified numerically that this relation is fulfilled
(again in the limit of $\Delta\to0$). Equation~(\ref{gl24}) is a special
case of a formula in the Appendix B of \cite{jar}.

The power $\beta$ of the algebraic part of Eq.~(\ref{gl21}) is found
to lie between
2 and 3. The fitted decay functions for $\gamma=1.8$ with the parameters
of Tab.~\ref{tab1}
--- which are useful representations of the data --- are given in
Fig.~\ref{fig3} as full lines. One notes that with $\Delta\to0$ the onset of
the algebraic decay is shifted to larger $L$ and to decreasing probability
level.

\section{Decay of the corresponding quantum system}
\label{sec3}
The almost exponential decay found above is in contrast to the behaviour of
the corresponding quantum system studied with the help of microwaves in
\cite{Alt1}. Suppose that the
holes through which the system is coupled to the external world (the antennas)
are small compared to the wave lengths occurring inside. Every hole may then be
identified with one decay channel c. Let $\langle\Gamma_c\rangle$ be the
decay width into channel c averaged over the eigenstates of the stadium. We
suppose that the decay widths into different channels are uncorrelated. To
avoid complications which are not instructive, we take $\langle\Gamma_c\rangle$
to be independent of c. The probability $P_q$ that the quantum system decays
at time $t$ after its formation is then
\begin{equation}
P_q(t)\sim(1+2\langle\Gamma_c\rangle t)^{-2-\frac{M}{2}},
\label{gl31}
\end{equation}
for $t$ larger than the equilibration time of the system,
see Sec.~(6.3) of \cite{HDM} or Eq.~(6) of \cite{Alt1}. Here, $M$ is the
number of open channels (or holes). The algebraic form
of Eq.~(\ref{gl31}) occurs because $P_q$ is an average over exponentially
decaying resonances whose decay amplitudes have a Gaussian distribution.
(See the introduction of \cite{DHM}).

The quantum system decays almost exponentially if $M$ is
large because Eq.~(\ref{gl31})
can then be approximated by
\begin{equation}
P_q(t)\sim \exp{[-(4+M)\langle\Gamma_c\rangle t]}
\label{gl32}
\end{equation}
and the weight of the algebraic tail becomes negligible.
For the channels to be statistically independent, any two antennas must,
however, be separated by a typical wave length or more.

We conclude that the quantum mechanical counterpart of a system with a small
hole decays algebraically. If the statistics
of the decay amplitudes is not exactly Gaussian it will decay
essentially algebraically.
Slight deviations
from Gaussian distribution have been detected \cite{KKS} and are due to the
presence of bouncing ball orbits between the parallel straight sections of
the stadium billiard.

Hence, the essentially algebraic decay of the quantum system has no classical
counterpart. The classical system with a small hole decays almost
exponentially.

What then is the origin of the algebraic tail in the classical decay
functions $P(L)$ ? --- It is the fact that the motion in the stadium is ergodic
only if it persists indefinitely. If the motion terminates by the escape of
the particle, it becomes apparent that ergodicity is not established in any
finite time. The reason for this is the existence of
marginally stable orbits that can almost trap the
particle --- the bouncing ball orbits. For details see the next section.

It is then the same class of orbits --- the marginally stable orbits --- that
add corrections to the essentially different behaviour of the classical and
the quantum systems.

\section{Model of delayed classical decay}
\label{sec4}
The argument justifying exponential decay with the decay constant $\lambda_0$
of Eq.~(\ref{gl22}) applies if ergodicity is established ``sufficiently
quickly''. A close inspection \cite{LeSo3} reveals that the fluctuations of
the frequency of the particles' arrival at location of the hole must be
small: Let
$\Delta t$ be the difference between two successive times of arrival at the
hole. Then $\overline{(\Delta t)^2}/\overline{\Delta t}^2$ must be small
compared to unity. Now, when there is a region of phase space into which
the particles
penetrate very slowly and --- by the same token --- in which they remain
trapped for an exceptionally long time once they are there,
then the above fluctuations are large and
the release of the trapped
particles will eventually dominate the decay process. Such regions of phase
space have been described e.g. in \cite{ViCa,Koo,PIK,FeSa,LeSo1,db} and the
first Ref. of \cite{Bunim1}: Close to the family
of bouncing ball orbits and the ``whispering gallery orbits'' there are
parts of phase space with volume $>0$ in which the particles can be trapped
for an arbitrarily long time. (A whispering gallery orbit is the motion
of the particle along the boundary).

The arguments of \cite{ViCa} show that the probability $G(n)$ for the particle
to be in an ``almost bouncing ball orbit'' that will persist for  more
than n collisions with the boundary is
\begin{equation}
G(n)=\frac{\gamma^2}{(\pi+2\gamma)n} ,\ \  n\gg1.
\label{gl41}
\end{equation}
An ``almost bouncing ball orbit'' allows for the angle between orbit and
boundary to be slightly different from $\pi/2$. In this situation, n
collisions amount to an orbit of length $L=2n$. (See the definition of
a ``collision with the boundary'' in
Sec. \ref{sec2}). Therefore
\begin{equation}
G(n(L))=\frac{1}{\alpha_0L} ,\ \ L\gg1 ,
\label{gl42}
\end{equation}
is the probability for the particle to be in an orbit that will
persist over a length $>L$.
Here, we have used
\begin{equation}
\alpha_0=\frac{\pi+2\gamma}{2\gamma^2}.
\label{gl43}
\end{equation}

The probability to be in a whispering gallery orbit decreases more strongly
than $L^{-1}$ and is therefore disregarded in the further discussion.

Hence, the probability that the particle is in an almost bouncing ball orbit
that will  persist for the length $L$ is asymptotically
\begin{equation}
g_{as}(L)=-\frac{d}{dL}G(n(L))=\frac{1}{\alpha_0L^2}.
\label{gl44}
\end{equation}
It was therefore anticipated in \cite{LeSo3,LeSo1} that $P(L)$ should
\mbox{asymptotically}
be $\sim L^{-2}$. There is some numerical evidence for this in \cite{PIK}:
Note that the function $N(t)$ given there by closed circles in
Fig.~2 is the present
$G(n(L))$. The present data confirm this asymptotic behaviour of $P(L)$ at
least qualitatively.

The following schematic model ---
inspired by the treatment of the Sinai billiard in \cite{FeSa} ---
shows how the algebraically delayed decay comes
about. Suppose that the phase space
can be split into two parts ${\cal C}$ and ${\cal L}$ such
that the decay happens in ${\cal C}$,
the delay in ${\cal L}$. Once  the particle is in ${\cal C}$, it
shall escape with probability
$\omega$ or immediately go back to ${\cal L}$ with
probability $1-\omega$. Consider
a particle which may be anywhere at the time $L=0$.
Define $g(L)dL$ as the probability  distribution for its next transition
from ${\cal L}$ into ${\cal C}$ to happen at time $L$. Then
the probability density $p_1(L)$ to escape at time $L$ after having made
exactly one transition from ${\cal L} \to {\cal C}$ is
\begin{equation}
p_1(L)=\omega g(L).
\label{gl45}
\end{equation}
Obviously $g_{as}$ in Eq.~(\ref{gl44}) is the asymptotic form of $g$.
Let $f(L)dL$ be the distribution of the time $L$ between two successive
transitions ${\cal L} \to {\cal C}$. Then the
probability density $p_2(L)$ for escape after exactly two
transitions ${\cal L} \to {\cal C}$ is
\begin{equation}
p_2(L)=\omega(1-\omega)g\otimes f,
\label{gl46}
\end{equation}
where the operator $\otimes$ denotes the convolution. For $k$ transitions
${\cal L} \to {\cal C}$ one has
\begin{equation}
p_k(L)=\omega(1-\omega)^{k-1}g\otimes(f\otimes\ldots f),
\label{gl47}
\end{equation}
where $g$ is folded with $(k-2)$ convolutions of $f$ with itself.
The decay probability $P(L)$ is the sum
\begin{equation}
P(L)=\sum_{k=1}^{\infty} p_k(L).
\label{gl48}
\end{equation}
Taking the Laplace transform
\begin{equation}
\hat{p}_k(s)=\int^\infty_0\,dL \exp{(-sL)}p_k(L)
\label{g49}
\end{equation}
converts the convolution into a product of the Laplace transforms $\hat{g}$ and
$\hat{f}$.
Therefore the transform $\hat{P}(s)$ of $P$ is a geometric series in
$\hat{f}$ which gives
\begin{equation}
\hat{P}(s)=\omega\hat{g}(s)[1-(1-\omega)\hat{f}(s)]^{-1}.
\label{gl410}
\end{equation}

The function $g$ is by definition proportional to an integral over $f$. We note
that this is numerically exhibited by \cite{Koo}, where the function $P_{fpt}$
corresponds to $f$ and its asymptotic behaviour is found to be close to
$\sim L^{-3}$. The relation between $g$ and $f$ is also just the relation
between ``transient chaos'' and ``chaotic scattering'' established in
\cite{PIK}. As the terms are used there, transient chaos corresponds to placing
at time zero the particle at random in phase space; chaotic scattering
corresponds to the situation in which the particle enters $\cal{L}$ at time
zero. The results of \cite{PIK} are compatible with $f\sim dg/dL$ as
it should be.

The normalization of $f$ and $g$ requires
\begin{equation}
\frac{dg}{dL}=-g(0)f(L)
\label{gl11}
\end{equation}
and the definition of the Laplace transform then leads to
\begin{equation}
\hat{f}(s)=1-s\frac{\hat{g}(s)}{g(0)}.
\label{gl412}
\end{equation}
Inversion of the Laplace transform (Eq.~(\ref{gl410})) then gives
$P(L)$ in the form of the Fourier integral
\begin{eqnarray}
P(L)=(2\pi)^{-1}\int^\infty_{-\infty}\,ds'\hat{P}(is')\exp{(is'L)}\nonumber\\
    =(2\pi)^{-1}\int^\infty_{-\infty}\,ds'\frac{\hat{g}(is')}{1+\frac{1-\omega}
     {\omega\alpha_0}is'\hat{g}(is')}\exp{(is'L)}.
\label{gl413}
\end{eqnarray}
A similar formula was used in Eq.~(5) of \cite{db} to describe the
exponential part only of the decay and to define the ``relaxation time''
within the closed stadium. It is shown here that Eq.~(\ref{gl413})
accounts for the decay function altogether if the appropriate probability
density $g(L)$ is introduced. It must necessarily allow for
large fluctuations of $L$.
There exists of course a $g(L)$ such that the observed decay
function is exactly reproduced. The present model is schematic by a naive
choice of $g(L)$ leaving open the exact definition of the partition of phase
space into $\cal{C}$ and $\cal{L}$. We only know the asymptotic form of $G$,
 see Eq.~(\ref{gl42}).
Let us --- in the simplest possible way --- assume that
\begin{equation}
g(L)=\alpha_0(1+\alpha_0 L)^{-2}.
\label{gl414}
\end{equation}
The decay function of eq.~(\ref{gl413}) approaches an exponential when
$\omega\to0$. The factor \mbox{$(1-\omega)/\omega\alpha_0$} in the
denominator of
the integrand is then very large and this makes the functional form of
$\hat{g}$ irrelevant. One can
thus approximately put \mbox{$\hat{g}(is')\approx\hat{g}(0)=1$} and the pole of
$\hat{P}(is')$ at \mbox{$s'\approx i\omega\alpha_0$}
yields \mbox{$P(L)\sim\exp{(-\omega\alpha_0 L)}$}.

The asymptotic behaviour of $P(L)$ is given by setting the denominator of
the integrand equal to unity. This is so because the omitted terms are of the
type $(is')^n\hat{g}(is')$ leading to derivatives of $g(L)$ which
decrease faster than $g(L)$. Hence, $P(L)\to g(L)$ for $L\to\infty$. This
means that the
asymptotic behaviour is independent of the size of the escape hole! This
result also implies $P(L)\sim L^{-2}$ for large $L$ as expected. Hence, the
present model incorporates the features of the data that we have termed
almost exponential decay.

By equating $\omega\alpha_0$ with the decay constant $\lambda_0$, Eqs.~
(\ref{gl22}) and (\ref{gl43}) determine the two parameters of the present
model.
This yields the theoretical decay functions given as dashed lines on
Fig.~\ref{fig3}
together with the data.

\section{Discussion}
\label{sec5}
The numerical experiment yields an almost exponential decay of the
classical Bunimovich stadium. This means: The decay function $P(L)$ starts out
exponentially and later turns to an algebraic behaviour. This contrasts with
the claim of \cite{BaBe} (see also the discussion in \cite{LeSo2}). The
weight of the algebraic decay, however, tends to zero with decreasing size
of the escape hole. In this modified sense of exponential decay we agree
with \cite{BaBe}.

Hillermeier et al. \cite{HBS} show how the algebraic decay can be understood
formally. The authors of \cite{KAR,MO1} have pointed out that the slow
transport of particles at the boundary of islands of regular motion can cause
algebraic decay. Yet there are no strictly stable orbits in the Bunimovich
stadium \cite{Bunim1}; in this sense it is a system with ``fully developed
chaos'' \cite{BOG}. However, there is the family of marginally stable bouncing
ball orbits which according to \cite{ViCa} causes the algebraic damping of
phase space correlations. It is responsible for the algebraic tail of the
decay function $P(L)$, too. More precisely, the bouncing ball orbits are  a
set of parabolic \cite{BER}, non-isolated periodic orbits. A similar set
of periodic orbits exists in the Sinai billiard and causes similar effects
\cite{FeSa}.

The algebraic tail of $P(L)$ and its suppression in the limit of vanishing
size $\Delta$ of the hole --- briefly the almost exponential decay --- is
semiquantitatively reproduced by the model inspired by \cite{FeSa}
and described in
Sec.~\ref{sec4}. The model reproduces the exponential part of $P(L)$.
It should yield a lower limit to the algebraic part because there may be
sources of delayed decay other than the bouncing ball orbits.
The comparison between data and model in Fig.~\ref{fig3}
shows this to be true.
Furthermore the data and the results of
the model should converge for $L\to\infty$ since the bouncing
ball orbits cause the most pronounced delay. Convergence is indeed indicated
by the curves for the largest holes although it is very slow; it
takes many inverse decay constants
$\lambda_0^{-1}$ before it is reached. One expects \cite{ViCa} that $P(L)$
tends --- for
$L\to\infty$ --- towards a function $g_{as}\sim L^{-2}$ which is independent
of the size of the hole and given by the properties of the bouncing ball orbits
alone. The model of
Sec. \ref{sec4} complies with this expectation. Again it is also compatible
with the
data: The experimental $P(L)$ asymptotically behaves as $L^{-\beta}$ with
$\beta$ somewhat larger than 2 and thus indicate convergence towards
$g_{as}(L)$. For the largest hole ($\Delta=0.25$) the convergence is
essentially achieved within the range of $L$ that we have studied.
It is, however, achieved so slowly that for the other hole sizes
$\Delta$, even orbit lengths of
$L=10^4$ are insufficient to demonstrate it. The data are, however, fully
consistent with convergence towards $g_{as}(L)$.

There are numerous studies of algebraically delayed {\it decay of correlations}
(as e.g. velocity correlations) in closed chaotic systems, see e.g.
\cite{Bunim1,ViCa,Mac,ZaGe}. The authors trace the delay back to the
existence of marginally stable periodic orbits. {\it Anomalous diffusion}
is caused by the same type of orbits, see e.g. \cite{GT,ArCa}. Again
these orbits are responsible for the algebraically delayed {\it escape}
of particles from fully chaotic systems. This emerges from
the arguments of \cite{LeSo1} and the numerical experiments of \cite{FeSa}
and the present note. Hence, all these phenomena are related to each other.
By using the considerations of \cite{ViCa} to define $g_{as}(L)$
we have directly linked the {\it decay of correlations} to the {\it escape}.

Furthermore, we have pointed out that the algebraic decay of chaotic quantum
systems is a consequence of wave mechanics and is not produced by the
delayed decay of the classical counterpart.

\acknowledgements
We thank C.H. Lewenkopf for very helpful discussions in the early stage of this
work and F. Neumeyer and S. Strauch for their assistance with the numerical
calculations. This work has been supported
by the Sonderforschungsbereich 185 ``Nichtlineare Dynamik'' of the Deutsche
Forschungsgemeinschaft and in part by the Bundesministerium f\"ur Forschung
und Technologie under contract number 06DA665I.
One of us (A.R.) is also grateful to the Institute for Nuclear Theory of the
University of Washington in Seattle, where this work was started, for its
hospitality and the Department of Energy for partial support.

\begin {figure}
\caption{Decay probability of the stadium with $\gamma=1.8$ and an escape hole
size of $\Delta=0.05$. The histogram is the result of the numerical experiment.
The full line is its parametrization via Eq.~(\ref{gl21}).
The parameters are given in Tab.~(\ref{tab1}).
The dashed line
is produced by the model of Sec.~\ref{sec4}.}
\label{fig1}
\end {figure}

\begin {figure}
\caption{The weight $A$ of the algebraic decay --- see Eq.~(\ref{gl21}) ---
vs. the size $\Delta$ of the escape hole. The errors are statistical ones.
They are given if larger than the size of the dots.}

\label{fig2}
\end{figure}

\begin {figure}
\caption{Decay probability $P$ vs. orbit length (or time) $L$ with
hole size $\Delta$ as parameter. The full lines show the parametrized
numerical experiment. The dashed lines result from the schematic model
described in the main text. }
\label{fig3}
\end{figure}

\begin{table}
\caption{Results of parametrizing the numerical experiments on the stadiums
with $\gamma=1.8$ and $\gamma=1$. The quantities listed are those of
Eqs.~(\ref{gl21}) and (\ref{gl22}) together with the normalized
$\chi^2$-values.}
\begin{tabular}{ccccccc}
$\gamma=1.8:$ &   &   &      &   & \\
$\Delta$&$A$&$\alpha[10^{-3}]$&$\beta$&$\lambda[10^{-3}]$&$\lambda_0[10^{-3}]$
&$\chi^2$\\
 \hline
0.2500& 0.301  & 34.00 & 2.36 & 34.93 & 30.77 & 0.99 \\
0.0500& 0.181  & 4.32 & 2.89 & 6.14 & 6.16 & 0.61\\
0.0100& 0.038 & 4.71 & 1.97 & 1.22  & 1.23 & 0.50\\
0.0025& 0 & / & / & 0.32  & 0.31& 0.82\\
       &   &   &      &   & \\
$\gamma=1:$   &   &   &      &   & \\
$\Delta$&$A$&$\alpha[10^{-3}]$&$\beta$&$\lambda[10^{-3}]$&$\lambda_0[10^{-3}]$
&$\chi^2$\\
 \hline
0.2500& 0.223 & 48.66 & 2.42 & 49.93  & 44.54 &1.38\\
0.0500& 0.128   & 6.40 & 2.96 & 8.88  & 8.91 &1.10\\
0.0100& 0.030 & 2.09 & 2.81 & 1.79    & 1.78&1.03\\
0.0025& 0  & /  & / & 0.47 & 0.46&1.01\\
\end{tabular}
\label{tab1}
\end{table}

\begin{references}
\bibitem {Alt1} H. Alt, H.-D. Gr\"af, H.L. Harney, R. Hofferbert,
                H.~Lengeler, A. Richter, P. Schardt and\\
                \mbox{H.A. Weidenm\"uller ,}
                Phys. Rev. Lett. {\bf 74} (1995) 62.
\bibitem {Bunim1} L.A. Bunimovich, Funct. Anal. Appl. {\bf 8} (1974) 254;
                  Commun. Math. Phys. {\bf 65} (1979) 295;
                  Sov. Phys. JETP {\bf 62} (1985) 842.
\bibitem {HBS} C.F. Hillermeier, R. Bl\"umel and U. Smilansky, Phys. Rev. A
               {\bf 45} (1992) 3486.
\bibitem {BaBe} W. Bauer and B. Bertsch, Phys. Rev. Lett. {\bf 65} (1990) 2213.
\bibitem {jar} C. Jarzynski, Phys. Rev. E {\bf 48} (1993) 4340.
\bibitem {HDM} H.L. Harney, F.-M. Dittes and A. M\"uller, Ann. Phys. (NY)
               {\bf 220} (1992) 159.
\bibitem {DHM} F.-M. Dittes, H.L. Harney and A. M\"uller, Phys. Rev. A
              {\bf 45} (1992) 701.
\bibitem {KKS} A. Kudrolli, V. Kidambi and S. Sridhar, Phys. Rev. Lett.
              {\bf 75} (1995) 822.
\bibitem {LeSo3} F.Mortessagne, O. Legrand and D. Sornette, Europhys. Lett.
                 {\bf 20} (1992) 287.
\bibitem {ViCa} F. Vivaldi, G. Casati and I. Guarneri, Phys. Rev. Lett.
                {\bf51} (1983) 727.
\bibitem {Koo} Koo-Chul Lee, Phys. Rev. Lett. {\bf 60} (1988) 1991.
\bibitem {PIK} A.S. Pikovsky, J. Phys. A {\bf 25} (1992) L477.
\bibitem {FeSa} A.J. Fendrik, A.M.F. Rivas and M.J. S\'anchez, Phys. Rev. E
                {\bf 50} (1994) 1948; A.J. Fendrik and M.J.S\'anchez,
                Phys. Rev. E {\bf 51} (1995) 2996.
\bibitem {LeSo1} O. Legrand and D. Sornette, Physica D {\bf 44} (1990) 229.
\bibitem {db} R.S. Dumont and P. Brumer, Chem. Phys. Lett. {\bf 188} (1992)
565.
\bibitem {LeSo2} O. Legrand and D. Sornette, Phys. Rev. Lett. {\bf 66} (1991)
                 2172.
\bibitem {KAR} C.F.F. Karney, Physica D {\bf 8} (1983) 360.
\bibitem {MO1} J.D. Meiss and E. Ott, Phys. Rev. Lett. {\bf 55} (1985) 2741;
               Physica D {\bf 20} (1986) 387.
\bibitem {BOG} S. Bleher, E. Ott and C. Grebogi, Phys. Rev. Lett. {\bf 63}
               (1989) 919.
\bibitem {BER} M.V. Berry in {\it Topics in Nonlinear Dynamics} (La Jolla
               Institute) Proc. Workshop on Topics in Nonlinear Dynamics,
               edited by S. Jorna,
               AIP - Conf. Proc. No.46 (AIP, NY 1978) p.16.
\bibitem {Mac}  J. Machta, J. Stat. Phys. {\bf 32} (1983) 555.
\bibitem {ZaGe} A. Zacherl, T. Geisel, J. Nierwetberg and G. Radons,
                Phys. Lett. A {\bf 114} (1986) 317.
\bibitem {GT} T. Geisel and S. Thomae, Phys. Rev. Lett. {\bf 52} (1984) 1936.
\bibitem {ArCa} R. Artuso, G. Casati and R. Lombardi, Phys. Rev. Lett
                {\bf 71} (1993) 62.

\end{references}
\end{document}